\documentstyle[aas2pp4]{article}
\input epsf
\begin{document}
 
\rightline{IASSNS-AST-97/6}
 
\title{Tensor Anisotropies in an Open Universe}
\author{Wayne Hu\footnote{Alfred P. Sloan Fellow}}
\affil{Institute for Advanced Study\\
School of Natural Sciences\\
Princeton, NJ 08540}
\authoremail{whu@sns.ias.edu}
\and
\author{Martin White}
\affil{Enrico Fermi Institute\\
University of Chicago\\
Chicago, IL 60637}
\authoremail{white@oddjob.uchicago.edu}
  
\begin{abstract}
\noindent
\rightskip=0pt
We calculate the anisotropies in the cosmic microwave background
induced by long-wavelength primordial gravitational waves in a universe with
negative spatial curvature, such as are produced in the ``open inflation''
scenario.  The impact of these results on the {\sl COBE\/} normalization of
open models is discussed.
\end{abstract}
 
\keywords{cosmology:theory -- cosmic microwave background}

There is considerable observational prejudice suggesting that we live in a
universe with negative spatial curvature, an open universe.  Recently
Bucher, Goldhaber and Turok (\cite{BGT}) have devised an open inflationary
cosmogony.  This has allowed one, for the first time, to calculate the
spectrum of primordial fluctuations in an open model.
The scalar field modes, which give rise to density perturbations,
come in 3 types and have been
extensively discussed in the literature
(for a recent review see Cohn~\cite{jdcReview}).
As with all inflationary models, a nearly scale-invariant spectrum of
gravitational waves (tensor modes) is also produced, and the spectrum of
such modes has recently been computed
(Tanaka \& Sasaki \cite{TanSas}, Bucher \& Cohn \cite{BucCoh}).
If the energy density during inflation is high enough, these modes induce
a measurable cosmic microwave background (CMB) anisotropy
(Abbott \& Wise \cite{AbbWis}).  
In this Letter we calculate these CMB anisotropies, and their implications
for the {\sl COBE\/} normalization of open models.  

\begin{figure*}[t]
\begin{center}
\leavevmode
\epsfxsize=6in \epsfbox{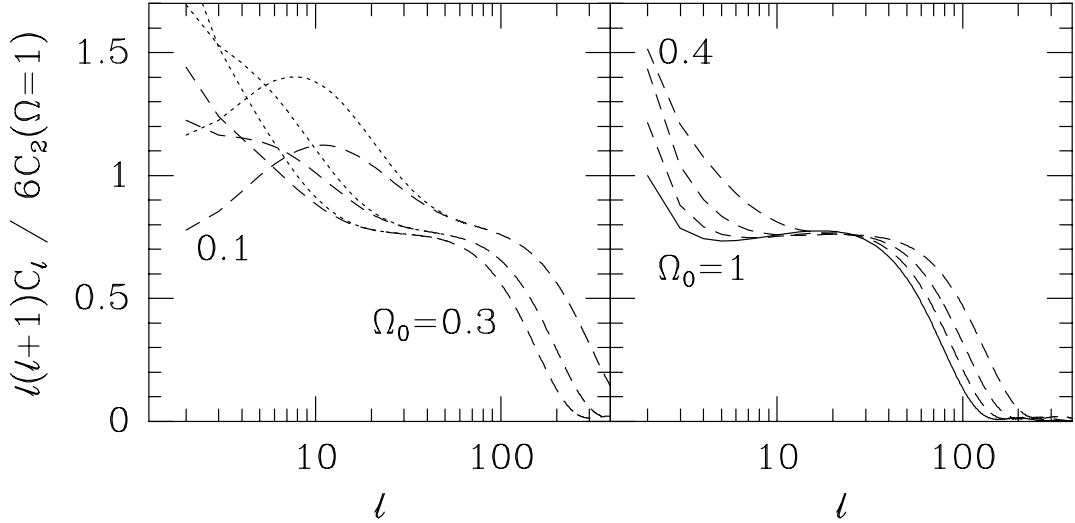}
\end{center}
\caption{The CMB anisotropy spectrum induced by a spectrum of gravitational
waves with $n_T=0$.  The solid line is $\Omega_0=1$ with $h=0.75$.
Dashed lines show the {\it minimal\/} anisotropy for (left) $\Omega_0=0.1$,
0.2 and 0.3, (right) 0.4, 0.6, 0.8 relative to $\Omega_0=1$.
The dotted lines (left) illustrate how low-$\ell$ anisotropies
are enhanced as the bubble parameters are altered from their
minimal values. }
\label{fig:tcl}
\end{figure*}

The gravitational waves represent tensor perturbations to the metric
\begin{equation}
ds^2 = a^2(\eta)\left[ -d\eta^2 +
  \left(\gamma_{ij}+h_{ij}\right)dx^i\, dx^j\right] \, ,
\label{eqn:tensormetric}
\end{equation}
where $a(\eta)$ is the scale factor, $d\eta=dt/a(t)$ is the conformal time
and $\gamma_{ij}$ is the 3-metric for a space of constant (negative) curvature $K = -H_0^2(1-\Omega_0)$.
The perturbations decompose as $h_{ij} = 2 h Q_{ij}$, 
where the harmonic modes are the transverse-traceless tensor eigenfunctions
of the Laplacian $\nabla^2 Q_{ij} = -k^2 Q_{ij}$
(Kodama \& Sasaki \cite{KodSas}, Abbott \& Schaefer \cite{AbbSch}).
The photon temperature distribution function $\Theta$, can likewise
be expanded in mode functions
$\Theta(\eta,\vec{x},\hat{n})=\sum_{\vec{k}}\sum_\ell
 \Theta_{\ell}(\vec{k})G_\ell(\vec{x},\vec{k},\hat{n}),$
which form a complete basis constructed out of covariant derivatives of
$Q_{ij}$.  

The Einstein equations reduce to a single relation which expresses the
evolution of the amplitude of the tensor metric perturbation in the
presence of tensor anisotropic stress in the matter $p\pi$
(Kodama \& Sasaki \cite{KodSas}, Abbott \& Schaefer \cite{AbbSch}),
\begin{equation}
\ddot h + 2{\dot a \over a} \dot h +
(k^2 + 2K)h = 8\pi G a^2 p \pi .
\label{eqn:eom}
\end{equation}
The quadrupolar variations in the metric induced by $\dot{h}$ leave a
corresponding signature through the photon quadrupole $(\ell=2)$ which acts
as the source to the tensor Boltzmann hierarchy
(Hu \& White \cite{TAMM})
\begin{eqnarray}
\dot \Theta_2 &=&
    -k {\sqrt{5}\over 7}\kappa_3 \Theta_3 - \dot h -
        {9 \over 10}\dot\tau\Theta_2 \,,\nonumber\\
\dot \Theta_\ell
&=& k\Bigg[ {\sqrt{\ell^2 -4 \vphantom{)}}
                \over (2\ell-1)}
        \kappa_\ell\Theta_{\ell-1}
                -
             {\sqrt{(\ell + 1)^2-4}
                \over (2\ell+3)}
        \kappa_{\ell+1} \Theta_{\ell+1} \Bigg]
        - \dot\tau \Theta_\ell\, , \qquad
\label{eqn:tenboltz}
\end{eqnarray}
where $\dot\tau=an_e\sigma_T$ is the differential optical depth to Thomson
scattering and the geodesic deviation factors are
$\kappa^2_\ell = [1 - (\ell^2 -3)K/k^2 ].$
In the above we have neglected the coupling of the temperature anisotropy to
the CMB polarization, since this has a small effect on the temperature
anisotropy spectrum calculated (see Hu \& White \cite{TAMM} and
its generalization to open universes Hu et al. \cite{Open} 
for more details).
The power spectrum of temperature anisotropies today is defined as,
\begin{equation}
(2\ell+1)^2 C_\ell^{(T)} = {2 \over \pi}
	\int_{0}^\infty {dq \over q}\ q^3
\ |\Theta_\ell(\eta_0,q)|^2 \, ,
\end{equation}
where $q^2=k^2+3K$ ranges from 0 to $\infty$ since the subcurvature modes
are complete for $k\ge \sqrt{-3K}$
(Abbott \& Schaeffer \cite{AbbSch}).  There are no supercurvature modes
for the gravity wave background (Tanaka \& Sasaki \cite{TanSas}).

The initial gravitational wave power spectrum may be parameterized
as
\begin{equation}
  q^3P_h(q) \equiv q^3 |h(\eta=0,q)|^2 \propto
  {(q^2+4)\over (q^2+1)}\, f(q) \, .
\end{equation}
The bubble models predict $f(q)\sim q$ for $q \ll 1$ and $f(q)\simeq1$ for
$q\ga 2$ (Tanaka \& Sasaki \cite{TanSas}, Bucher \& Cohn \cite{BucCoh}).
The {\it minimal\/} anisotropies are produced when this turnover occurs at
the largest allowed $q$, or $f(q)=\tanh(\pi q/2)$
(see Bucher \& Cohn \cite{BucCoh} Eqn. 6.5),
which we display in Fig.~\ref{fig:tcl} (dashed lines).
Decreasing the turnover scale increases the low-$\ell$ anisotropy
as shown in the $f(q)=\tanh(\pi q)$ example of 
Fig.~\ref{fig:tcl} (dotted lines).

The exact form of the function $f(q)$ is sensitive
to the parameters of the bubble.  However, as illustrated 
in Fig.~\ref{fig:tcl},
variations are confined to the low multipoles for
reasonable values of $\Omega_0\ga0.3$, as was the case for the scalar modes
(Yamamoto \& Bunn~\cite{YamBun}).  Small changes in bubble parameters
will thus be lost in cosmic variance.
Since a calculation of non-scale invariant spectra has not been performed
for tensors, we make an ansatz that the spectral tilt is a pure power-law
in $k$, viz.~$(k/H_0)^{n_T}$ times the above result.

\begin{figure}[t]
\begin{center}
\leavevmode
\epsfxsize=4.5in \epsfbox{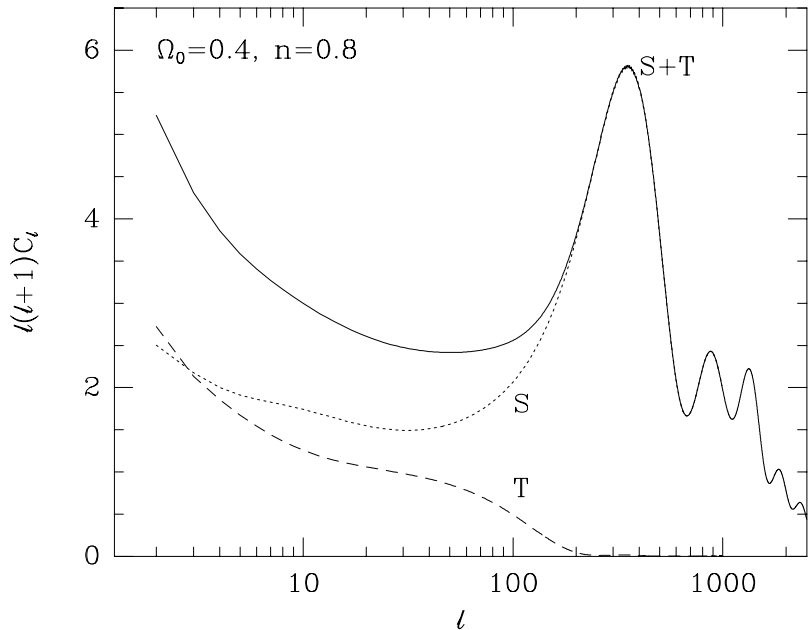}
\end{center}
\caption{Relative tensor and scalar anisotropy contributions for a model
with $\Omega_0=0.4$ and $n=0.8$ assuming $n_S-1=n_T$,
$\Omega_{\rm B}h^2=0.02$ and $h=0.6$.  This model has been chosen for
illustration to have $T/S\simeq 1$ and 
would produce insufficient large-scale
structure, e.g.~$\sigma_8\simeq0.25$ as well as 
provide a poor fit to the
{\sl COBE\/} data.}
\label{fig:total}
\end{figure}

In flat space, as $k\to 0$ the power in the Boltzmann hierarchy of
Eqn.~(\ref{eqn:tenboltz}) remains in the {\it quadrupole}.
Thus we expect that tensor contributions in a flat universe will have an
enhanced quadrupole due to long wavelength modes
(as seen in Fig.~\ref{fig:tcl}).  However as $k\to 0$, Eqn.~(\ref{eqn:eom})
requires that $\dot{h}\to 0$, so the source of the anisotropy dies off
to low $k$, ensuring a finite quadrupole.  
If $K<0$ then even as $q\to 0$, the metric perturbation damps when
$\eta\ga |K|^{-1/2}$, i.e.~when the curvature scale crosses the horizon.
This provides a source to the anisotropy to arbitrarily low $q$, and indeed
the $C_\ell$ from a scale invariant spectrum, $q^3 P(q)=$ const., would
diverge logarithmically.
A similar divergence would occur for the scalar monopole due to the decay of
the gravitational potential
(see Fig.~\ref{fig:total} and Sugiyama \& Silk \cite{SugSil})
but the monopole is not observable.
The open inflation models regulate this divergence with the finite energy
difference before and after the tunnelling event which defines the bubble
(Tanaka \& Sasaki \cite{TanSas}, Bucher \& Cohn \cite{BucCoh}), 
and translates into the turnover in $f(q)$ discussed above.
If one goes to sufficiently low $\Omega_0\la 0.1$, the effect of the curvature
cutoff $k(q=0)=\sqrt{-3K}$ can also be seen as a low multipole suppression in
the spectrum.  However for $\Omega_0$ of interest for structure formation the
curvature cutoff is absent and the large contribution from low-$q$ is the
dominant effect, leading to anisotropy spectra which decrease strongly with
$\ell$ on {\sl COBE\/} scales.

Since gravitational waves provide anisotropies but no density fluctuations,
they lower the normalization of the matter power spectrum.  Open models
already have quite a low normalization
(e.g.~White \& Silk \cite{WhiSil}, White \& Scott \cite{WhiSco})
so we seek the {\it minimal\/} anisotropies induced by gravity waves.
These minimal anisotropies are also in fact close to what most models would
predict, with reasonable inflationary potentials.
We express the {\sl COBE\/} 4-year normalization (Bennett et al.~\cite{Ben})
in terms of the value of the density perturbations per logarithmic interval
in $k$ evaluated at horizon crossing.  Writing $\Delta^2(k)=k^3P(k)/(2\pi^2)$
we define $\delta_{\rm H}\equiv\Delta(k=H_0)$.
If we hold the ``shape'' of the matter power spectrum fixed, the small scale
power (e.g.~$\sigma_8$) is proportional to $\delta_{\rm H}$.
For minimal models with $\widetilde{n}\equiv n-1=n_T$ the {\sl COBE\/} 4-year
data give
\begin{equation}
10^5\ \delta_{\rm H}=1.95\; \Omega_0^{-0.35-0.19\ln\Omega_0+0.15\widetilde{n}}
  \exp\left[ 1.02 \widetilde{n} + 1.70 \widetilde{n}^2 \right] \, .
\label{eqn:cobe}
\end{equation}
The fitting function works to 3\% over the range $0.2<\Omega_0\le1$ 
and
$0.7<n<1$ whereas the {\sl COBE\/} $1\sigma$ error is 10\%.
Compared with the equivalent expression {\it without\/} gravity waves
(Bunn \& White~\cite{BunWhi} Eqn. 31), we isolate the additional
suppression as 
$\Omega_0^{0.32 \widetilde{n}} \exp(2.02\widetilde{n} 
+ 1.84\widetilde{n}^2)$.
This suppression exacerbates the problem that low-$\Omega_0$ models have with
the present day abundance of rich clusters (White \& Silk \cite{WhiSil}).

In conclusion we have presented the first calculation of the anisotropy in
the CMB from a spectrum of long-wavelength gravitational waves in an open
universe.  The spectrum exhibits a peak in large-angle power which
is dependent on the modifications to the initial power spectrum 
near the curvature scale from the bubble wall.  
For very low $\Omega_0$ a curvature cut-off is seen in the spectrum, analogous
to the case of scalar modes.
Since gravitational waves provide anisotropies but no density fluctuations,
they lower the normalization of the matter power spectrum.
We have calculated this normalization from the {\sl COBE\/} 4-year data and
expressed our result in terms of a fitting function, Eqn.~(\ref{eqn:cobe}).
The lower normalization of the tilted models with gravitational waves
exacerbates the difficultly such models have in fitting the present day
abundance of rich clusters.

\bigskip
We would like to thank Joanne Cohn for several useful conversations
and Douglas Scott for comments on a draft of this work.  W.H. was 
supported by the W.M. Keck Foundation.

\end{document}